\documentclass[a4paper,10pt]{article}

\usepackage{bm,amsmath,amssymb,graphicx,mathrsfs}

\newcommand{\nc}{\newcommand}
\nc{\rnc}{\renewcommand}
\nc{\nn}{\nonumber}
\nc{\bra}{\langle}
\nc{\ket}{\rangle}
\rnc{\(}{\left(}
\rnc{\)}{\right)}
\rnc{\[}{\left[}
\rnc{\]}{\right]}
\DeclareMathOperator{\Tr}{Tr}

\begin{document}

\title{The matrix product representation for the $q$-VBS state of
one-dimensional higher integer spin model}

\author{Kohei Motegi\thanks{E-mail: motegi@gokutan.c.u-tokyo.ac.jp} \\
\it Okayama Institute for Quantum Physics, \\
\it Kyoyama 1-9-1,
Okayama 700-0015, Japan}

\date{\today}

\maketitle

\begin{abstract}
The generalized $q$-deformed valence-bond-solid groundstate
of one-dimensional higher integer spin model is studied.
The Schwinger boson representation and the matrix product
representation of the exact groundstate is determined, which
recovers the former results for the spin-1 case or the isotropic limit.
As an application, several correlation functions are
evaluated from the matrix product representation.
\end{abstract}

\section{Introduction}
In one-dimensional quantum systems, a completely different behavior 
for the integer spin
chains from the half-integer spin chains was predicted the Haldane
\cite{Haldane1,Haldane2}.
The antiferromagnetic isotopric
spin-1 model introduced by Affleck, Kennedy, Lieb and Tasaki
(AKLT model) \cite{AKLT},
whose groundstate can be exactly calculated, has been a useful toymodel
for the deep understanding of Haldane's prediction of the massive behavior
for integer spin chains, such as the
discovery of the special type of long-range order \cite{DR,Tasaki}. 

The AKLT model has been generalized to
higher spin models, anisotropic models, etc \cite{AAH,KSZ1,KSZ2,KSZ3,Osh,TS1,BY,GR,SR,TZX,TZXLN,KM,AHQZ}.
The hamiltonians are essentially linear combinations
of projection operators with nonnegative coefficients.

In this paper we consider the anisotropic integer spin-$S$ Hamiltonian
\begin{align}
&H=\sum_{k=1}^L H(k,k+1), \\
&H(k,k+1)=\sum_{J=S+1}^{2S}C_J(k,k+1) \pi_{J}(k,k+1),
\end{align}
where $C_J(k,k+1) \ge 0$, and $\pi_{J}(k,k+1)$, which acts
on the $k$-th and $(k+1)$-th site,
is the $U_q(su(2))$ projection operator
for $V_{S} \otimes V_{S}$ to $V_J$ where $V_j$ is the $(2j+1)$-dimensional
representation of the quantum group $U_q(su(2))$ \cite{Drinfeld,Jimbo}.
We determine the matrix product representation for the groundstate,
which is useful for calculations of correlation functions.
For $S=1$ or $q=1$ limit, it recovers the known results
for the isotropic spin-$S$ model or anisotropic spin-1 model
\cite{KSZ2,KSZ3,TS1,TS2}.
Several correlation functions are evaluated from the matrix product representation.

This paper is organized as follows. In the next section, we
briefly review the quantum group $U_q(su(2))$.
By use of the Weyl representation of $U_q(su(2))$, we 
construct a boson representation for the valence-bond-solid (VBS) groundstate.
The matrix product representation for the VBS state is constructed in section 3,
from which several correlation
functions are evaluated for $S=2$ and $S=3$. Section 4 is devoted to conclusion.
\section{Schwinger boson representation of the groundstate}
The quantum group $U_q(su(2))$ is defined by generators $X^+, X^-,H$
with relations
\begin{align}
{[} X^+, X^- {]}=\frac{q^H-q^{-H}}{q-q^{-1}},  \ \ [H, X^{\pm}]=\pm 2 X^{\pm}.
\end{align}
The comultiplication is given by
\begin{align}
\Delta(X^+)&=X^+ \otimes q^{H/2}+q^{-H/2} \otimes X^+, \\
\Delta(X^-)&=X^- \otimes q^{H/2}+q^{-H/2} \otimes X^-, \\
\Delta (H)&=H \otimes 1+1 \otimes H.
\end{align}
For convenience, let us define $q$-integer, $q$-factorial and $q$-binomial coefficients
as
\begin{align}
[n]_q=\frac{q^n-q^{-n}}{q-q^{-1}}, \ [n]_q!=\prod_{k=1}^n [k], \
\left[
\begin{array}{c}
n \\
k
\end{array}
\right]_q
=\frac{[n]_q!}{[k]_q![n-k]_q!}.
\end{align}
$U_q(su(2))$ has the Schwinger boson representation 
\cite{qboson1,qboson2,qboson3}.
Introducing two $q$-bosons $a$ and $b$ satisfying
\begin{align}
&aa^{\dagger}-q a^{\dagger} a=q^{-N_a}, \ \  bb^{\dagger}-q b^{\dagger} b=q^{-N_b}, \\
&[N_a, a]=-a, \ \ [N_a, a^{\dagger}]=a^{\dagger}, \ \
[N_b, b]=-b, \ \ [N_b, b^{\dagger}]=b^{\dagger},
\end{align}
$U_q(su(2))$ can be realized through the relations
\begin{align}
X^+=a^{\dagger} b, \ \ X^-=b^{\dagger} a, \ \ H=N_a-N_b.
\end{align}
The basis of $(2j+1)$-dimensional representation $V_j$ is given by
\begin{align}
|j,m \ket=\frac{(a^{\dagger})^{j+m}(b^{\dagger})^{j-m}}{([j+m]_q! [j-m]_q!)^{1/2}}
|\mathrm{vac} \ket, \ \ (m=-j, \dots, j).
\end{align}
We construct the VBS groundstate in terms of Schwinger bosons, following
the arguments of \cite{KX}.
Let us denote the $q$-bosons $a$ and $b$ acting on the $l$-th site as $a_l$ and $b_l$.
We utilize the Weyl representation of $U_q(su(2))$ 
\cite{weyl1,weyl2} for convenience.
$a_l^{\dagger}$ and $b_l^{\dagger}$ is represented as multiplication by
variables $x_l$ and $y_l$ on the space of polynomials $\mathbb{C}[x_l,y_l]$,
respectively.
$a_l$ and $b_l$ are represented as difference operators
\begin{align}
a_l=\frac{1}{(q-q^{-1})x_l}(D_q^{x_l}-D_{q^{-1}}^{x_l}), \ \
b_l=\frac{1}{(q-q^{-1})y_l}(D_q^{y_l}-D_{q^{-1}}^{y_l}),
\end{align}
where
\begin{align}
D_{p}^{x_l}f(x_l,y_l)=f(px_l,y_l), \ \ D_{p}^{y_l}f(x_l,y_l)=f(x_l,py_l).
\end{align}
Then, at the $l$-th site, one has
\begin{align}
X_l^{+}=\frac{x_l}{(q-q^{-1})y_l}(D_q^{y_l}-D_{q^{-1}}^{y_l}), \ \
X_l^{-}=\frac{y_l}{(q-q^{-1})x_l}(D_q^{x_l}-D_{q^{-1}}^{x_l}), \ \
q^{H_l}=D_q^{x_l} D_{q^{-1}}^{y_l}.
\end{align}
The basis of $(2S_l+1)$-dimensional representation $V_{S_{l}}$ is given by
\begin{align}
\{x_l^{S_l+m_l} y_l^{S_l-m_l} \ | \ m_l=-S_l, \dots S_l \}.
\end{align}
The tensor product of two irreducible representations has the following
Clebsch-Gordan decomposition
\begin{align}
V_{S_k} \otimes V_{S_l}= \oplus_{J=|S_k-S_l|}^{S_k+S_l}V_J.
\end{align}
The highest weight vector $v_J \in V_J$ has the following form
\begin{align}
v_J=\sum_{m_k+m_l=J}C_{m_k,m_l}
x_k^{S_k+m_k}y_k^{S_k-m_k} x_l^{S_l+m_l}y_l^{S_l-m_l}.
\end{align}
Since
\begin{align}
X_{kl}^+ v_J=& \Delta(X_{kl}^+) \sum_{m_k+m_l=J}C_{m_k,m_l}
x_k^{S_k+m_k}y_k^{S_k-m_k} x_l^{S_l+m_l}y_l^{S_l-m_l} \nn \\
=&\sum_{m_k=0}^{J-1}
([S_k-m_k]_q q^{J-m_k}C_{m_k,J-m_k}+[S_l-J+m_k+1]_q q^{-m_k-1}
C_{m_k+1,J-m_k-1}) \nn \\
&\times x_k^{S_k+m_k+1}y_k^{S_k-m_k-1} x_l^{S_l+J-m_k}y_l^{S_l-J+m_k},
\end{align}
one has
\begin{align}
C_{m_k,J-m_k}=\frac{(-1)^{S_k-m_k}
\left[
\begin{array}{c}
S_k+S_l-J \\
S_k-m_k
\end{array}
\right]_q
}{
(-1)^{S_k}
\left[
\begin{array}{c}
S_k+S_l-J \\
S_k
\end{array}
\right]_q
}
q^{m_k(J+1)}C_{0,J}. \label{relation1}
\end{align}
Utilizing \eqref{relation1} and
\begin{align}
\prod_{j=1}^m(1-zq^{2j-2})
=\sum_{k=0}^m (-z)^k q^{k(m-1)}
\left[
\begin{array}{c}
m \\
k
\end{array}
\right]_q,
\end{align}
one gets
\begin{align}
v_J=\frac{q^{S_k(J+1)}C_{0,J}}
{(-1)^{S_k}
\left[
\begin{array}{c}
S_k+S_l-J \\
S_k
\end{array}
\right]_q
}
x_k^{S_k-S_l+J}x_l^{S_l-S_k+J}
\prod_{m=1}^{S_k+S_l-J}(x_k y_l-q^{2m-2-S_k-S_l}x_l y_k).
\end{align}
We are now considering the homogeneous chain, i.e.,  $S_k=S$ for all $k$.
The highest weight vector $v_S \in V_S \subset V_S \otimes V_S$
is divisible by
$\prod_{m=1}^S(q^m x_k y_l-q^{-m} y_k x_l)$. Moreover, we conjecture the following. \\
\\
{\bf Conjecture} \\
All vectors in $V_j \subset V_S \otimes V_S, \ j=0,1, \dots, S$ are divisible by 
$\prod_{m=1}^S(q^m x_k y_l-q^{-m} y_k x_l)$. \\
\\
We have checked this conjecture for several values of $S$.
The vectors for the case $S=2$ are listed in the Appendix.
Based on this conjecture and the property of
projection opreators $\pi_J w_K=\delta_{JK}w_K, w_K \in V_K$
, we have the $q$-deformed lemma of Lemma 1 in
\cite{KX}. \\
\\
{\bf Lemma} \\
All solutions of
\begin{align}
\pi_J(k,k+1)| \psi \ket=0, \ \ S+1 \le J \le 2S,
\end{align}
for fixed $k$ can be represented in the following form
\begin{align}
| \psi \ket=f(a_k^{\dagger}, b_k^{\dagger},a_{k+1}^{\dagger}, b_{k+1}^{\dagger})
\prod_{m=1}^S(q^m a_k^{\dagger} b_{k+1}^{\dagger}-q^{-m} b_k^{\dagger} a_{k+1}^{\dagger})
| \mathrm{vac} \ket,
\end{align}
where $f(a_k^{\dagger}, b_k^{\dagger},a_{k+1}^{\dagger}, b_{k+1}^{\dagger})$ is
some polynomial in 
$a_k^{\dagger}, b_k^{\dagger},a_{k+1}^{\dagger}$ and $b_{k+1}^{\dagger}$. \\
\\
From this Lemma, we find the $q$-deformed VBS groundstate is
\begin{align}
| \Psi \ket_{PBC}=\prod_{k=1}^L \prod_{m=1}^S
(q^m a_k^{\dagger} b_{k+1}^{\dagger}-q^{-m} b_k^{\dagger} a_{k+1}^{\dagger})
|\mathrm{vac} \ket,
\end{align}
where $a_{L+1}=a_1, b_{L+1}=b_1$ for the periodic chain, and
\begin{align}
| \Psi \ket_{p_1,p_2}=Q_{\mathrm{left}}(a_1^{\dagger},b_1^{\dagger};p_1)
\prod_{k=1}^{L-1} \prod_{m=1}^S
(q^m a_k^{\dagger} b_{k+1}^{\dagger}-q^{-m} b_k^{\dagger} a_{k+1}^{\dagger})
Q_{\mathrm{right}}(a_L^{\dagger},b_L^{\dagger};p_2)
|\mathrm{vac} \ket,
\end{align}
where
\begin{align}
Q_{\mathrm{left}}(a_1^{\dagger},b_1^{\dagger};p_1)&=
\left[
\begin{array}{c}
S \\
p_1-1
\end{array}
\right]_q^{1/2}
(a_1^{\dagger})^{S-p_1+1} (b_1^{\dagger})^{p_1-1}, \ \ (p_1=1, \dots S+1), \\
Q_{\mathrm{right}}(a_L^{\dagger},b_L^{\dagger};p_2)&=
\left[
\begin{array}{c}
S \\
p_2-1
\end{array}
\right]_q^{1/2}
(a_L^{\dagger})^{p_2-1} (b_L^{\dagger})^{S-p_2+1}, \ \ (p_2=1, \dots S+1),
\end{align}
for the open chain, generalizing the results of \cite{AAH}.
\section{Matrix product representation}
In the last section, we constructed the $q$-VBS states in terms of
Schwinger bosons.
One can transform them in the matrix product representation
as in \cite{TS1,TS2}, which are
\begin{align}
| \Psi \ket_{PBC}&=\Tr [g_1 \otimes g_2 \otimes \cdots  \otimes g_{L-1} \otimes g_L], \\
| \Psi \ket_{p_1,p_2}&=[g^{\mathrm{start}} \otimes g_2 \otimes \cdots g_{L-1} \otimes 
g_L]_{p_1,p_2},
\end{align}
where $g_k$ and $g^{\mathrm{start}}$ are $(S+1) \times (S+1)$ matrices whose 
matrix elements are given by
\begin{align}
g_k(i,j)=&(-1)^{S-i+1} q^{(2i-2-S)(S+1)/2} \nn \\
&\times
\left(
\left[
\begin{array}{c}
S \\
i-1
\end{array}
\right]_q
\left[
\begin{array}{c}
S \\
j-1
\end{array}
\right]_q
\right)^{1/2}
(a_k^{\dagger})^{S-i+j} (b_k^{\dagger})^{S+i-j} | \mathrm{vac} \ket_k \nn \\
=&(-1)^{S-i+1} q^{(2i-2-S)(S+1)/2} \nn \\
&\times
\left(
\left[
\begin{array}{c}
S \\
i-1
\end{array}
\right]_q
\left[
\begin{array}{c}
S \\
j-1
\end{array}
\right]_q
[S-i+j]_q! [S+i-j]_q!
\right)^{1/2}
|S;j-i \ket_k, \\
g^{\mathrm{start}}(i,j)=
&
\left(
\left[
\begin{array}{c}
S \\
i-1
\end{array}
\right]_q
\left[
\begin{array}{c}
S \\
j-1
\end{array}
\right]_q
[S-i+j]_q! [S+i-j]_q!
\right)^{1/2}
|S;j-i \ket_k.
\end{align}
For $q \to 1$ limit, one recovers the results of \cite{TS2}.
We can also construct the matrix product representation in the following form
\begin{align}
| \Psi \ket_{PBC}&=\Tr [f_1 \otimes f_2 \otimes \cdots  \otimes f_{L-1} \otimes f_L],
\end{align}
where
\begin{align}
f_k(i,j)=&(-1)^{S-i+1} q^{(i+j-2-S)(S+1)/2} \nn \\
&\times
\left(
\left[
\begin{array}{c}
S \\
i-1
\end{array}
\right]_q
\left[
\begin{array}{c}
S \\
j-1
\end{array}
\right]_q
[S-i+j]_q! [S+i-j]_q!
\right)^{1/2}
|S;j-i \ket_k, 
\end{align}
which reproduces the result for $S=1$ \cite{KSZ2,KSZ3}.

From the matrix product representation, one can formulate correlation functions.
Let $f_j^{\dagger}$ be a matrix replacing the ket vectors of the matrix $f_j$ 
by the bra vectors.
We define $(S+1)^2 \times (S+1)^2$ matrices $G$ and $G^A$ as
\begin{align}
G_{(m_{j-1},n_{j-1};m_j,n_j)}&=f_j^{\dagger}(m_{j-1},m_j)f_j(n_{j-1},n_j), \\
G^A_{(m_{j-1},n_{j-1};m_j,n_j)}&=f_j^{\dagger}(m_{j-1},m_j)A_jf_j(n_{j-1},n_j).
\end{align}
Explicitly we have
\begin{align}
G_{(a,b;c,d)}=&\delta_{a-b,c-d}(-1)^{a+b}q^{(a+b+c+d-2S-4)(S+1)/2} \nn \\
&\times \left(
\left[
\begin{array}{c}
S \\
a-1
\end{array}
\right]_q
\left[
\begin{array}{c}
S \\
b-1
\end{array}
\right]_q
\left[
\begin{array}{c}
S \\
c-1
\end{array}
\right]_q
\left[
\begin{array}{c}
S \\
d-1
\end{array}
\right]_q
\right)^{1/2} \nn \\
&\times
([S-a+c]_q![S+a-c]_q![S-b+d]_q![S+b-d]_q!)^{1/2}.
\end{align}
The eigenvalues of $G$ for $S=2$ are
\begin{align}
&\lambda_1=[5]_q[4]_q[2]_q, \\
&\lambda_2=\lambda_3=\lambda_4=-[5]_q[2]_q^2, \\
&\lambda_5=\lambda_6=\lambda_7=\lambda_8=\lambda_9=[2]_q^2.
\end{align}
Moreover, we conjecture that the eigenvalues of $G$ for general $S$ is given by
\begin{align}
\lambda(l)=(-1)^l \frac{[2S+1]_q!}{[S+1]_q} 
\frac{
\left[
\begin{array}{c}
S \\
l
\end{array}
\right]_q
}
{
\left[
\begin{array}{c}
S+l+1 \\
l
\end{array}
\right]_q
}, \ (l=0,1,\dots,S),
\end{align}
where the degeneracy of $\lambda(l)$ is $2l+1$. \\
For $A=S^z$, one has
\begin{align}
G^{S^z}_{(a,b;c,d)}=&\delta_{a-b,c-d}(d-b)(-1)^{a+b}q^{(a+b+c+d-2S-4)(S+1)/2} \nn \\
&\times \left(
\left[
\begin{array}{c}
S \\
a-1
\end{array}
\right]_q
\left[
\begin{array}{c}
S \\
b-1
\end{array}
\right]_q
\left[
\begin{array}{c}
S \\
c-1
\end{array}
\right]_q
\left[
\begin{array}{c}
S \\
d-1
\end{array}
\right]_q
\right)^{1/2} \nn \\
&\times
([S-a+c]_q![S+a-c]_q![S-b+d]_q![S+b-d]_q!)^{1/2}.
\end{align}
One point function $\bra A \ket$
and two point function $\bra A_1 B_r \ket$
of the periodic chain can be represented as
\begin{align}
\bra A \ket&=(\Tr G^L)^{-1} \Tr G^A G^{L-1}, \label{onepoint} \\
\bra A_1 B_r \ket&=(\Tr G^L)^{-1} \Tr G^A G^{r-2} G^B G^{L-r}. \label{twopoint}
\end{align}
Denoting the eigenvalues and the normalized eigenvectors of $G$ as
$|\lambda_1| > |\lambda_2| \ge \cdots \ge |\lambda_{(S+1)^2}|$ and 
$| e_1 \ket, |e_2 \ket, \dots |e_{(S+1)^2} \ket$,
\eqref{onepoint} and \eqref{twopoint} reduces to
\begin{align}
\bra A \ket&=\lambda_1^{-1} \bra e_1|G^A|e_1 \ket, \\
\bra A_1 B_r \ket&=\sum_{n=1}^{(S+1)^2} \lambda_n^{-2} 
\left( \frac{\lambda_n}{\lambda_1} \right)^r
\bra e_1|G^A|e_1 \ket \bra e_n|G^B|e_1 \ket.
\end{align}
in the thermodynamic limit $L \to \infty$.

Let us calculate several correlation functions. For $S=2$,
the probability of finding $S^z=m$ value $\bra P(S^z=m) \ket$ is
\begin{align}
&\bra P(S^z=2) \ket=\bra P(S^z=-2) \ket=\frac{1}{[5]_q}, \\
&\bra P(S^z=1) \ket=\bra P(S^z=-1) \ket=\frac{[2]_q[8]_q}{[5]_q[4]_q^2}, \\
&\bra P(S^z=0) \ket=\frac{[2]_q}{[5]_q[4]_q} \left(1+\frac{[12]_q}{[3]_q[4]_q} \right).
\end{align}
In the $q=1$ limit, $\bra P(S^z=m) \ket=1/5$ for all $m$.
As we move away from $q=1$, $P(S^z=0)$ increases,
i.e., the spins prefer the transverse $x$-$y$ plane. \\
The  spin-spin correlation function $\bra S_1^z S_r^z \ket$ is
\begin{align}
\bra S_1^z S_r^z \ket=&-\frac{[2]_q[3]_q}{[4]_q}
\left( \frac{[2]_q}{[5]_q[4]_q} \right)^r
\Big\{
(q-q^{-1})(q^3-q^{-3})\frac{[6]_q^2}{[3]_q^2[2]_q^2}+[2]_q^2(-[5]_q)^r
\Big\},
\end{align}
which reduces to $-6(-2)^{-r}$ for $q=1$.
$\bra S_1^z S_r^z \ket$
exhibits exponential decay
for large distances, which is a typical behavior of gapful systems. 

For $S=3$, one has
\begin{align}
\bra S_1^z S_r^z \ket=&-\frac{[2]_q}{[6]_q[5]_q[3]_q} 
\left( \frac{[3]_q}{[7]_q[6]_q[5]_q} \right)^r
\Big\{ (q-q^{-1})^2(q^3-q^{-3})^2([9]_q-(q^2-q^{-2})^2)^2           
\frac{[4]_q^2}{[2]_q^2}(-[2]_q)^r                          \nn \\
&+(q^3-q^{-3})^2 \frac{[8]_q^2[5]_q}{[4]_q^2}([7]_q[2]_q)^r
+([2]_q^4-2 [3]_q)^2 \frac{[6]_q[2]_q}{[3]_q}(-[7]_q[6]_q)^r
\Big\},
\end{align}
which reduces to $-80 (-3)^{r-2} 5^{-r}$ in the $q=1$ limit.

\section{Conclusion}
In this paper, we considered one-dimensional
spin-$S$ $q$-deformed AKLT models.
We derived the Schwinger boson representation
and the matrix product representation for the valence-bond-solid groundstate.
The matrix product representation is practical for calculating
correlation functions. The spin-spin correlation functions exhibit exponential
decay for large distances.

An interesting problem is to calculate the entanglement entropy of this model,
which is a typical quantification of the entanglement of quantum systems.
It is interesting to see how the
entanglement entropy changes as we move away from the isotropic point 
\cite{FKR,KHH,XKHK} (see also \cite{KHK,KKKKT} for other VBS states).
\section*{Acknowledgement}
This work was partially supported
by Global COE Program (Global Center of Excellence for
Physical Sciences Frontier) from the Ministry of Education,
Culture, Sports, Science and Technology, Japan.

\section*{Appendix}
We list all the vectors in $v_j \in V_j \subset V_{S} \otimes V_{S}, j=1,2,\dots,S$. \\
$S=2$
\begin{align}
v_2 &\propto x_k^2 x_l^2(qx_k y_l-q^{-1}x_l y_k)(q^2x_k y_l-q^{-2}x_l y_k), \nn \\
(X_{kl}^-)v_2 &\propto x_k x_l(q^{-2}x_k y_l+q^2 x_l y_k)
(qx_k y_l-q^{-1}x_l y_k)(q^2x_k y_l-q^{-2}x_l y_k), \nn \\
(X_{kl}^-)^2 v_2 &\propto \{ q^{-4}x_k^2 y_l^2+(q+q^{-1})^2 
x_k x_l y_k y_l+q^4 x_l^2 y_k^2 \}
(qx_k y_l-q^{-1}x_l y_k)(q^2x_k y_l-q^{-2}x_l y_k), \nn \\
(X_{kl}^-)^3 v_2 &\propto y_k y_l(q^{-2}x_k y_l+q^2 x_l y_k)
(qx_k y_l-q^{-1}x_l y_k)(q^2x_k y_l-q^{-2}x_l y_k), \nn \\
(X_{kl}^-)^4 v_2 &\propto y_k^2 y_l^2(qx_k y_l-q^{-1}x_l y_k)(q^2x_k y_l-q^{-2}x_l y_k), 
\nn \\
v_1 &\propto x_k x_l(x_k y_l-x_l y_k)
(qx_k y_l-q^{-1}x_l y_k)(q^2x_k y_l-q^{-2}x_l y_k), \nn \\
(X_{kl}^-)v_1 &\propto (q^{-2}x_k y_l+q^2 x_l y_k)
(x_k y_l-x_l y_k) 
(qx_k y_l-q^{-1}x_l y_k)(q^2x_k y_l-q^{-2}x_l y_k), \nn \\
(X_{kl}^-)^2v_1 &\propto y_k y_l(x_k y_l-x_l y_k)
(qx_k y_l-q^{-1}x_l y_k)(q^2x_k y_l-q^{-2}x_l y_k), \nn \\
v_0 &\propto (q^{-1}x_k y_l-qx_l y_k)(x_k y_l-x_l y_k)
(qx_k y_l-q^{-1}x_l y_k)(q^2x_k y_l-q^{-2}x_l y_k). \nn
\end{align}

\end{document}